\documentclass[aps,preprint,amsmath,amssymb,showpacs]{revtex4}
\usepackage{graphicx}
\begin{document}

\title{Anomalous interactions of the top quark with photon via flavor changing neutral currents at the CLIC}

\author{A. A. Billur}
\email[]{abillur@cumhuriyet.edu.tr} \affiliation{Department of
Physics, Cumhuriyet University, 58140, Sivas, Turkey}

\author{M. K\"{o}ksal}
\email[]{mkoksal@cumhuriyet.edu.tr} \affiliation{Department of
Physics, Cumhuriyet University, 58140, Sivas, Turkey}

\begin{abstract}
Photon-photon reactions provide an excellent opportunity to isolate the $tq\gamma$ vertex.
For this purpose, we have examined the potential of the $e^{-}e^{+}\rightarrow e^{-}\gamma^{*}\gamma^{*}e^{+}\rightarrow e^{-}t\bar{q}e^{+}\rightarrow e^{-}W b\bar{q}e^{+}$ ($\gamma^{*}$ is the Weizsacker-Williams photon and $\bar{q}=\bar{u},\bar{c}$) process to investigate the anomalous $tq\gamma$ couplings in $\gamma^{*}\gamma^{*}$ collisions at the CLIC. We have obtained $95\%$ confidence level
limits on the anomalous couplings for various values of the center-of-mass energy and integrated luminosity.
We have shown that the limit on anomalous $\kappa_{tq\gamma}$ coupling is more restricted
with respect to current experimental limits.
\end{abstract}


\maketitle

\section{Introduction}
The top quark is the heaviest available fundamental particle in the
Standard Model (SM). Because of the large mass of the top quark,
it's interactions are an excellent probe of the electroweak
symmetry-breaking mechanism, and they should therefore play an important
role in the search of physics beyond the SM. For this purpose, particularly, the anomalous interactions of the top quark
can be examined by flavor changing neutral currents (FCNC).
In the SM, FCNC
decays $t\rightarrow q\gamma$ ($q=u,c$) cannot be observed at tree level, but these
decays can only make loop contributions. As a result, such processes
are anticipated to be enormously rare within the SM with branching
ratios of an order of $10^{-10}$ \cite{he,ee1,ee2}. However, various models beyond the SM such as the minimal supersymmetric model \cite{18,19,20,21,22,23,24}, two-Higgs doublet model \cite{8,9,at,10,11,12,13}, the quark-singlet model \cite{14,15,16}, extra dimension models \cite{25,26,go}, the littlest Higgs model \cite{at1,at2,sun2}, the topcolor-assisted technicolor model \cite{27,cao,yeni,zha} or supersymmetry \cite{17} could lead to a very large
increase of FCNC processes involving the top quark.

Present experimental constraints at $95\%$ confidence level (C. L.) on the anomalous $tq\gamma$ couplings
are obtained from two limits: ·$\kappa_{tu\gamma} < 0.12$ supplied by ZEUS collaboration \cite{35} and

\begin{eqnarray}
BR(t \rightarrow u
\gamma)+BR(t \rightarrow c
\gamma)<3.2\%
\end{eqnarray}
presented by CDF collaboration \cite{28}.

The FCNC anomalous interactions among the top quark, two quarks $u$, $c$ and
the photon can be written in a model independent way with dimension five effective
lagrangian as follows  \cite{35}

\begin{eqnarray}
\textit{L}= \sum_{q=u,c} g_{e} e_{t}\bar{t}\frac{i\sigma_{\mu\nu}p^{\nu}}{\Lambda}\kappa_{tq\gamma}q A^{\mu}
\end{eqnarray}
where $g_{e}$ is the electromagnetic coupling constant, $e_{t}$ is the top quark electric charge, $\kappa_{tq\gamma}$  denotes the strength of the anomalous couplings of top quark with photon, $\Lambda$ is an effective cut-off scale which is conventionally set to the mass of the top quark \cite{35},
$\sigma_{\mu\nu}=
(\gamma^{\mu}\gamma^{\nu}-\gamma^{\nu}\gamma^{\mu})/2$ with
$\gamma^{\mu}$ which stands for the Dirac matrix, and $p$ is the momentum of photon.

Also, using the interaction lagrangian in Eq.($2$), the anomalous
decay width of the top quark can be easily obtained as follows

\begin{eqnarray}
\Gamma(t \rightarrow
q\gamma)=\frac{g_{e}^2 e_{t}^{2}\kappa_{tq\gamma}^{2}m_{t}^{3}}{8 \pi \Lambda^{2}}         &&           (q=u,c)
\end{eqnarray}
where the masses of $u$ and $c$ quarks are omitted in the above
equation. Since the dominant decay mode of the top quark is $t\rightarrow b W$,
the branching ratio of anomalous $t\rightarrow q\gamma$ decay
generally is given by the following formula:

\begin{eqnarray}
BR(t\rightarrow q\gamma)=\frac{\Gamma(t\rightarrow
q\gamma)}{\Gamma (t\rightarrow b W)}.
\end{eqnarray}
Therefore, using the equations ($1$) and ($4$), we can obtain
the magnitude of the upper limits of anomalous coupling provided by CDF collaboration as follows

\begin{eqnarray}
\kappa_{tq\gamma}=0.29.
\end{eqnarray}

In the literature, the interactions of the top quark via FCNC have been experimentally and theoretically examined \cite{h1,h2,h3,h4,h44,h5,h6,h7,h8,h9,h99,h10,h11,h12,h13,ko1,ko2,ko3,han1,han2,sun1,aol1,29,aol2,30,31,32,33,34,36,37,38,koksal1}.

The LHC might allow us to observe the top quark's FCNC couplings with its high center-of-mass energy
and high luminosity. However, the signal which may occur from the new physics beyond the SM is difficult to determine due to jets coming from proton remnants at the LHC. On the other hand, a linear lepton collider, which has extremely high luminosity and a clean experimental environment, can provide complementary
information for these properties while carrying out high precision measurements that
would complete the LHC results. One of the proposed high luminosity and high energy lepton colliders is the Compact Linear Collider (CLIC) \cite{h14}, which is designed to be constructed
in three main stages as given in Table I \cite{CLIC1}.

The CLIC provides a suitable platform to examine the $e \gamma$ and $\gamma\gamma$ processes by converting the incoming leptons into an intense beam of high-energy photons. On the other hand, $\gamma^{*} e$ and $\gamma^{*}\gamma^{*}$ processes at the CLIC occur instinctively by the virtual photon emitted from the original $e^{-}$ or $e^{+}$ beam. Therefore, $\gamma^{*} e$ and $ \gamma^{*} \gamma^{*} $ processes are more realistic than $e \gamma$ and $\gamma\gamma$ processes occurring
through the Compton backscattering mechanism.
$\gamma^{*}\gamma^{*}$ processes have been described by the Weizsacker-Williams
approximation \cite{es1,es2,es3}. In this approximation, the emitted photons are scattered
at very small angles from the beam pipe. Since the emitted photons have a low
virtuality, they are almost real.

$\gamma^{*} \gamma^{*}$ processes can isolate the $tq\gamma$ couplings from $tqZ$ couplings. Jets which originate from light quarks ($u$, $d$, and $s$ quarks) differ from heavy quarks ($c$ and $b$ quarks) in the final state \cite{h5,ucexp1,ucexp2}.
Therefore, the anomalous $\kappa_{tu\gamma}$ coupling could be distinguished from $\kappa_{tc\gamma}$ coupling via the process $e^{-}e^{+}\rightarrow e^{-}\gamma^{*}\gamma^{*}e^{+}\rightarrow e^{-}t\bar{q}e^{+}\rightarrow e^{-}W b\bar{q}e^{+}$, if the anomalous coupling $\kappa_{tu\gamma}$ is not equal to $\kappa_{tc\gamma}$. So far, new physics research through $ \gamma^{*} \gamma^{*}$ processes  at
the LEP \cite{den1,den2}, Tevatron \cite{den3,den4,den5,den6,den7} and LHC \cite{den8,den9} have been experimentally studied in the literature.
As a result, the CLIC as a $\gamma^{*}\gamma^{*}$ collider provides us with an important opportunity to investigate the anomalous couplings  of the top quark.
A schematic diagram describing this process is represented in Fig.$1$.

\section{Sensitivity to Anomalous Couplings}

In the case of the effective lagrangian in Eq.$(2)$, the subprocess $\gamma^{*}\gamma^{*}  \rightarrow t \bar{q}$
is described by four tree-level diagrams (Fig.$2$). In our calculations, we perform the simulation program
COMPHEP-4.5.1 by including the new interaction vertices \cite{COM}. In addition, the acceptance cuts were used as the following

\begin{eqnarray}
p_{T}^{b,\bar{q}}>20 \: GeV,
\end{eqnarray}

\begin{eqnarray}
|\eta_{b,\bar{q}}|<2.5,
\end{eqnarray}

\begin{eqnarray}
\Delta R_{b,\bar{q}}>0.4.
\end{eqnarray}
where  $p_{T}$ are the transverse momentum cuts of the $b$ and $\bar{q}$ quarks, $\eta$ denotes pseudorapidity, and $\Delta R$ is the
separation of the $b$ and $\bar{q}$ quarks.

The integrated total cross-sections of the process $e^{-}e^{+}\rightarrow e^{-}\gamma^{*}\gamma^{*}e^{+}\rightarrow e^{-}W b\bar{q}e^{+}$ as a function of the anomalous couplings $\kappa_{tq\gamma}=\kappa$ for $\sqrt{s}=0.5, 1.5$ and $3$ TeV are plotted in Fig.$3$ for $Q^2_{max}=4$ GeV$^2$.
Anomalous ${tq\gamma}$ couplings are defined by effective
lagrangian in Eq.$(2)$, and it has an energy dimension of $5$. However, the total cross
section including the SM and new physics has a higher energy dependence than the SM cross
section. Therefore, we can see from this figure that the total cross section increases with increasing the center-of-mass energy.

On the other hand, there may occur an uncertainty arising from the virtuality of $\gamma^{*}$ used in the Weizsacker-Williams approximation. We have obtained SM cross sections for various photon virtualities as follows: $\sigma_{SM}=3.44\times10^{-5}$ pb at $\sqrt{s}=0.5$ TeV and $Q^2_{max}=4$  GeV$^2 $, $\sigma_{SM}=4.69\times10^{-5}$ pb at $\sqrt{s}=0.5$ TeV and $Q^2_{max}=64$ GeV$^2 $,  $\sigma_{SM}=2.29\times10^{-4}$ pb at $\sqrt{s}=1.5$ TeV and $Q^2_{max}=4$  GeV$^2 $,  $\sigma_{SM}=3 \times10^{-4}$ pb at $\sqrt{s}=1.5$ TeV and $Q^2_{max}=64$  GeV$^2 $,   $\sigma_{SM}=4.37\times10^{-4}$ pb at $\sqrt{s}=3$ TeV and $Q^2_{max}=4$  GeV$^2 $,  $\sigma_{SM}=5.78\times10^{-4}$ pb at $\sqrt{s}=3$ TeV and $Q^2_{max}=64$  GeV$^2 $ for the process $e^{-}e^{+}\rightarrow e^{-}\gamma^{*}\gamma^{*}e^{+}\rightarrow e^{-}W b\bar{q}e^{+}$. In Fig. $4$, in detail, the integrated cross sections are given for different $Q_{max}^{2}$ values as a function of $\kappa$. We can see from these figures the total cross section changes slightly with the variation of the $Q_{max}^{2}$ value. 

As shown in Fig. $5$, we plot the invariant mass distributions for the $Wb$ system in the final state, the signal has a peak over the SM background. In the SM, there is no single top quark production at tree level via the subprocess $\gamma^{*}\gamma^{*}\rightarrow t \bar{q}$.
Hence, the differential cross section depending on the invariant mass distribution of the $Wb$ system is a good observable for the top quark's FCNC couplings. In Fig.$6$, the rapidity distributions of the final state $b$ quark at the $\sqrt{s}=0.5, 1.5$ and $3$ TeV. It can be understood that the rapidity distributions of the final state $b$ quark from the new physics signal and SM background are located generally in the range of $|\eta^{b}|<2$. In Fig.$7$, the transverse momentum distributions of the final state $b$ quark at the $\sqrt{s}=0.5, 1.5$ and $3$ TeV. We can see from Fig.$ 7 $ that the transverse momentum of the final state $b$ quark from the new physics signal and SM background can be discerned at especially large transverse momentum values.

A comprehensive examination of the anomalous coupling $\kappa_{tq\gamma}$
requires a statistical analysis. In this work, we calculate
sensitivity of the $e^{-}e^{+}\rightarrow e^{-}\gamma^{*}\gamma^{*}e^{+}\rightarrow e^{-}W b\bar{q}e^{+}$ process to anomalous coupling $\kappa_{tq\gamma}$ using two different statistical analysis methods. First, we employ a one-parameter $\chi^{2}$ test when the number of SM events is greater than $10$.
The $\chi^{2}$ test without a systematic error is given by

\begin{eqnarray}
\chi^{2}=\left(\frac{\sigma_{SM}-\sigma_{AN}}{\sigma_{SM}\delta_{stat}}\right)^{2}
\end{eqnarray}
where $\sigma_{AN}$ is the total cross section including SM and new physics, $\delta_{stat}=\frac{1}{\sqrt{N_{SM}}}$: $N_{SM}$ is the
number of SM events calculated as $N_{SM}=\sigma_{SM}\times L_{int} \times \epsilon \times BR(W\rightarrow \ell \nu_{\ell})$.
Here $L_{int}$ is the integrated luminosity, $\epsilon=60\%$ is the efficiency for $b$-tagging. We take into account the leptonic decay of the
$W$ boson with the branching through $W\rightarrow\ell \nu_{\ell}$, where $\ell=e,\mu$. 

In the second analysis, we use a Poisson distribution, which is the appropriate
sensitivity analysis for the number of SM events smaller than or equal
to $10$. In Poisson analysis, the sensitivity limits are obtained assuming the number of observed events are equal
to the SM prediction, i.e.,

\begin{eqnarray}
N_{obs}=\sigma_{SM}\times L_{int} \times \epsilon \times BR(W\rightarrow \ell \nu_{\ell})
\end{eqnarray}

Upper limits of the number of events $N_{up}$ at $95\%$ C.L. can be obtained as follows \cite{ses3,ses4}

\begin{eqnarray}
\sum_{k=0}^{N_{obs}}P_{Poisson}(N_{up};k)=0.05.
\end{eqnarray}

The sensitivity limits on the anomalous coupling $\kappa$ for different values of photon virtuality, center-of-mass energy and luminosity have been given in Tables II-IV.  It has shown that the bounds on the anomalous coupling are slightly improved when $Q_{max}^{2}$ increases. We can also understand that the large values of $Q_{max}^{2}$ do not bring an important contribution to obtain sensitivity limits on the anomalous coupling at large energy and luminosity values. As shown in the tables, this process can develop the sensitivity limits of the anomalous coupling $\kappa$ with respect to the current experimental limits. 

\section{Conclusions}

Even though the LHC has high luminosity and high energy, it does not have clean environment due to the strong interactions. Therefore linear colliders, which have less background than the LHC, provide an opportunity for precise measurements of the top quark anomalous couplings.
Photon-photon collisions at the linear colliders enable us to investigate FCNC top quark interactions without a contribution coming from the anomalous $tqZ$ coupling. These processes may also be more efficient to distinguish $\kappa_{tu\gamma}$ and $\kappa_{tc\gamma}$ in comparison to hadron colliders due to the $u$ and $c$ quarks appearing in the final state.

In this paper we have analyzed the $e^{-}e^{+}\rightarrow e^{-}\gamma^{*}\gamma^{*}e^{+}\rightarrow e^{-}W b\bar{q}e^{+}$  process with the anomalous ${tq\gamma}$ couplings in a model independent effective lagrangian approach at the CLIC. So far, ZEUS and CDF collaborations have the most stringent limits on anomalous FCNC couplings of the photon with top quark. In our study, we present more restrictive limits with respect to the obtained anomalous couplings $\kappa_{tq\gamma}$ than the current experimental limits and these anomalous couplings depend strictly on the center-of-mass energy and integrated luminosity.
Consequently, the CLIC make an important contribution to the search for the anomalous ${tq\gamma}$ couplings through the $e^{-}e^{+}\rightarrow e^{-}\gamma^{*}\gamma^{*}e^{+}\rightarrow e^{-}W b\bar{q}e^{+}$ process.

\pagebreak

\pagebreak

\begin{figure}
\includegraphics{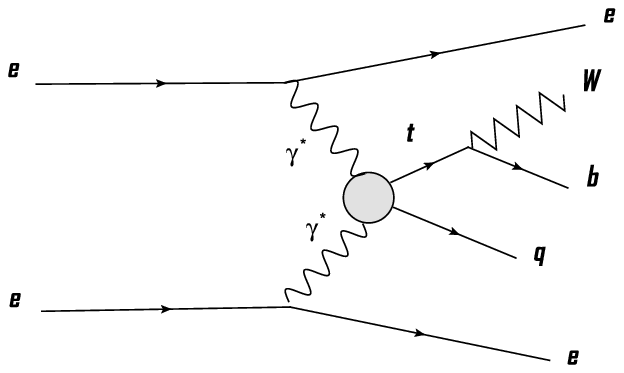}
\caption{Schematic diagram for the process $e^{-}e^{+}\rightarrow e^{-}\gamma^{*}\gamma^{*}e^{+}\rightarrow e^{-}t\bar{q}e^{+}\rightarrow e^{-}W b\bar{q}e^{+}$.
\label{fig1}}
\end{figure}

\begin{figure}
\includegraphics{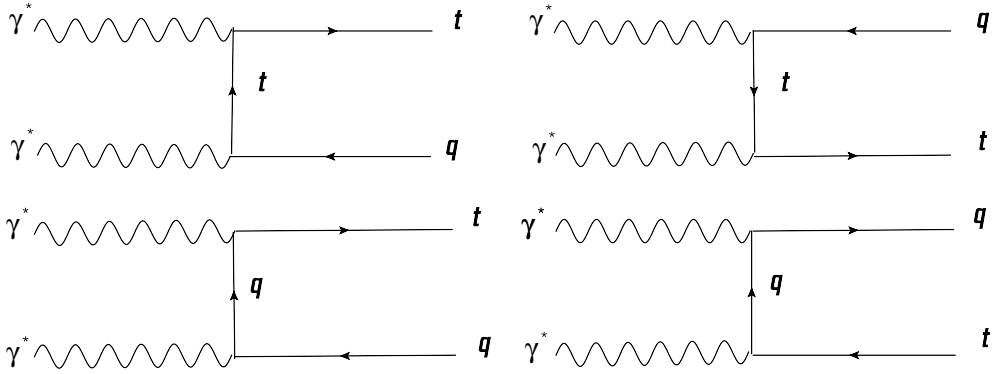}
\caption{Tree-level Feynman diagrams for the subprocess $\gamma^{*}\gamma^{*}\rightarrow t \bar{q}$ ($q=u,c$).
\label{fig2}}
\end{figure}

\begin{figure}
\includegraphics{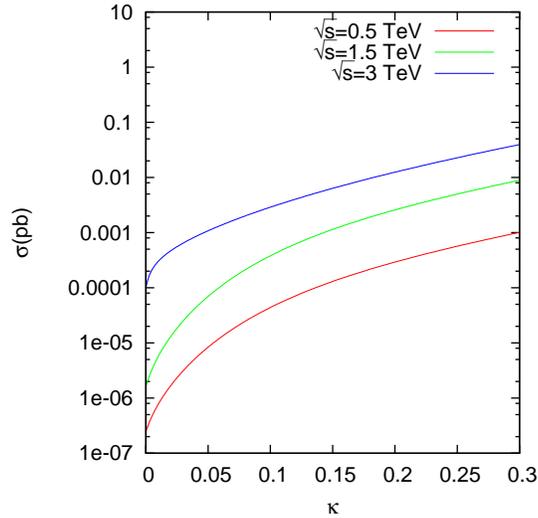}
\caption{For $Q^2_{max}=4$ GeV$^2$, the integrated total cross-sections of the process $e^{-}e^{+}\rightarrow e^{-}\gamma^{*}\gamma^{*}e^{+}\rightarrow e^{-}W b\bar{q}e^{+}$ as a
function of the anomalous coupling $\kappa_{tq\gamma}=\kappa$ for three different center-of-mass energies.
\label{fig3}}
\end{figure}

\begin{figure}
\includegraphics[width=1.0\textwidth]{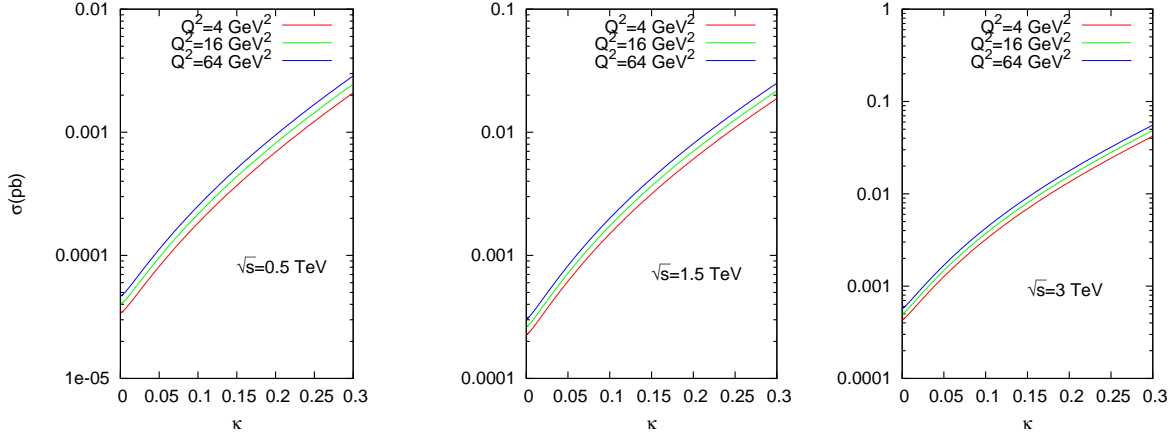}
\caption{The total cross sections as a function of the anomalous coupling $\kappa_{tq\gamma}=\kappa$ for three different center-of-mass energies and different values of $Q^{2}$ for the process $e^{-}e^{+}\rightarrow e^{-}\gamma^{*}\gamma^{*}e^{+}\rightarrow e^{-}W b\bar{q}e^{+}$.
\label{fig4}}
\end{figure}

\begin{figure}
\includegraphics[width=1.0\textwidth]{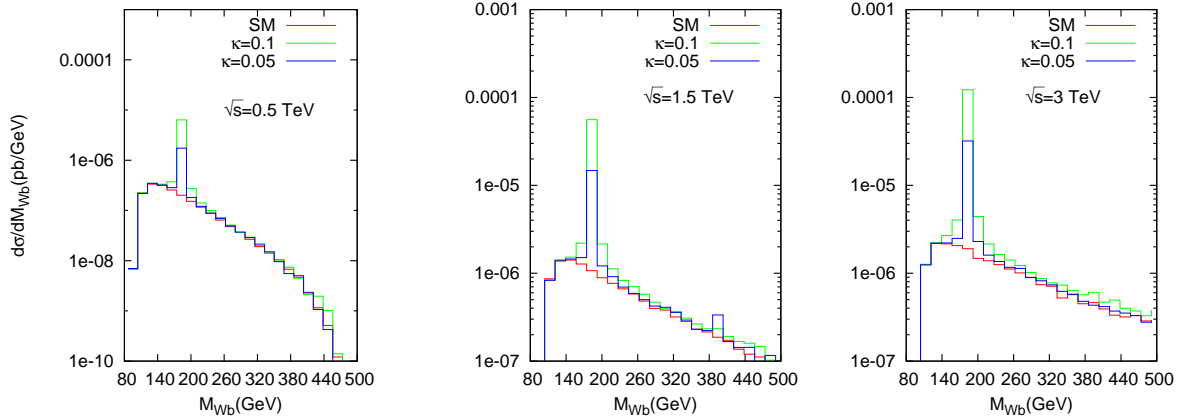}
\caption{The invariant mass distributions of the final state $Wb$ system of the process $e^{-}e^{+}\rightarrow e^{-}\gamma^{*}\gamma^{*}e^{+}\rightarrow e^{-}W b\bar{q}e^{+}$ for SM and signal with different anomalous coupling $\kappa$ values at $\sqrt{s}=0.5,1.5$ and $3$ TeV.
\label{fig5}}
\end{figure}

\begin{figure}
\includegraphics[width=1.0\textwidth]{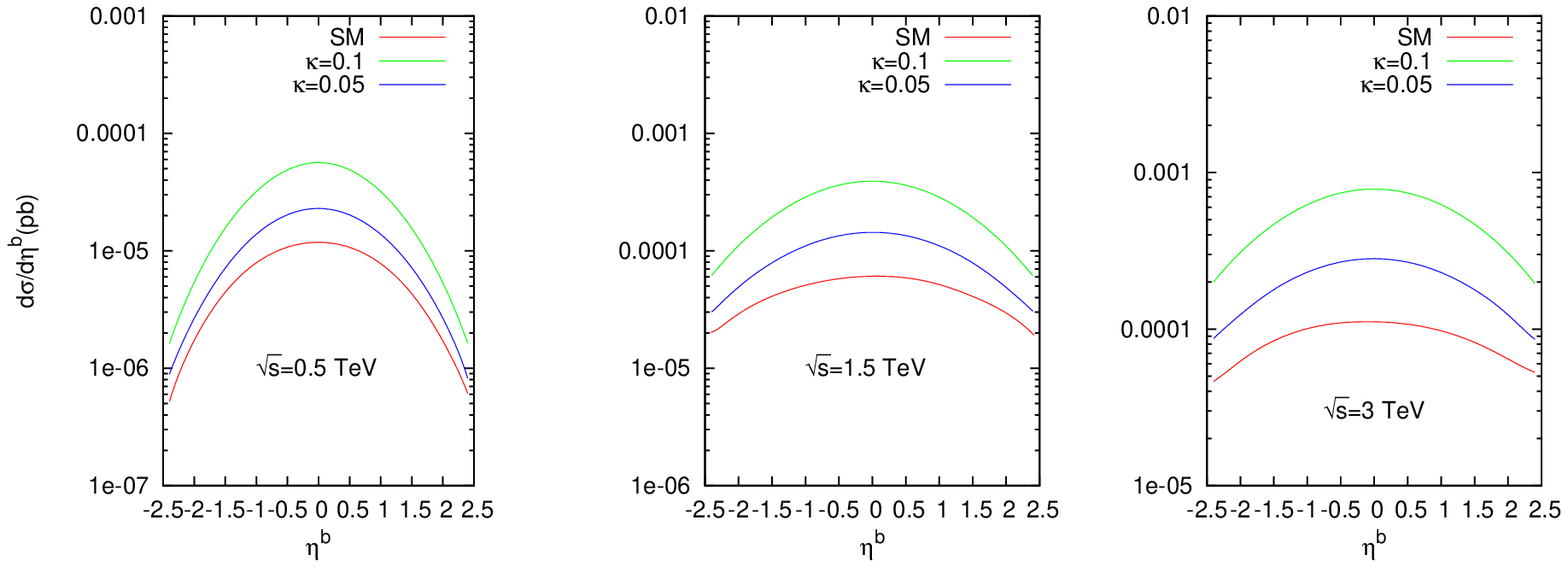}
\caption{The rapidity distributions of the final state b quark for the process $e^{-}e^{+}\rightarrow e^{-}\gamma^{*}\gamma^{*}e^{+}\rightarrow e^{-}W b\bar{q}e^{+}$ for SM and signal with different anomalous coupling $\kappa$ values at $\sqrt{s}=0.5,1.5$ and $3$ TeV.
\label{fig6}}
\end{figure}

\begin{figure}
\includegraphics[width=1.0\textwidth]{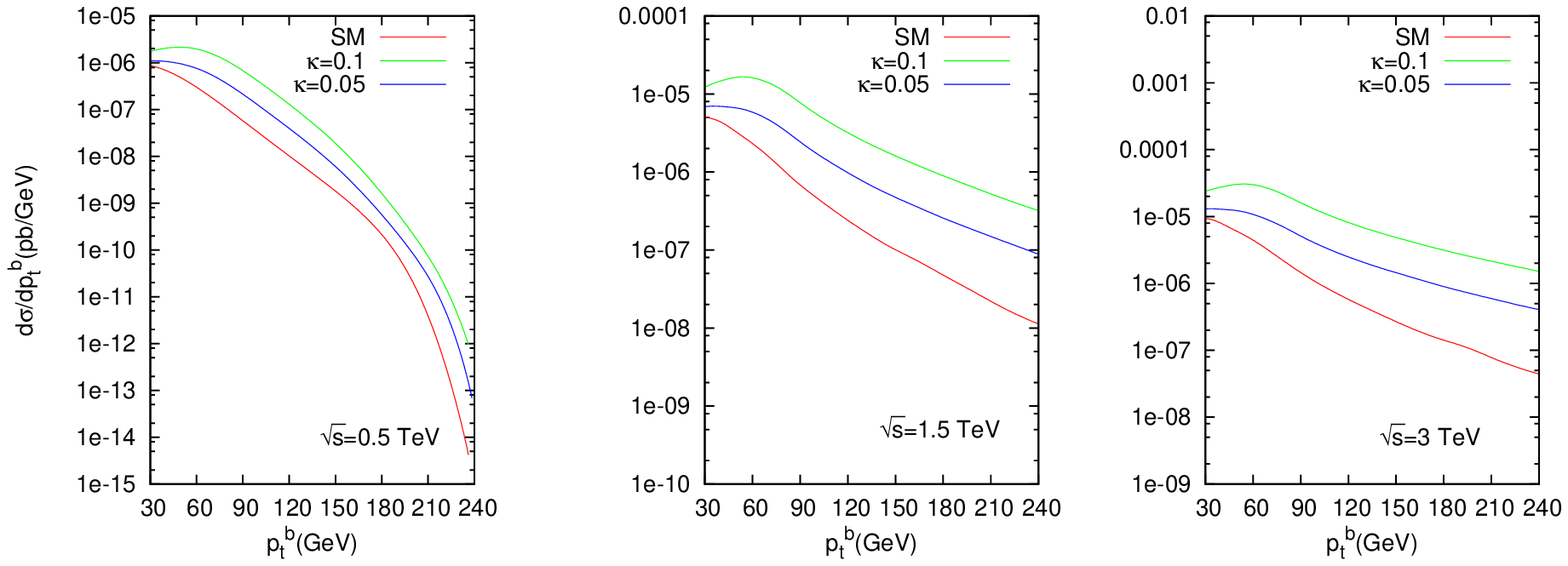}
\caption{The transverse momentum distributions of the final state b quark for the process $e^{-}e^{+}\rightarrow e^{-}\gamma^{*}\gamma^{*}e^{+}\rightarrow e^{-}W b\bar{q}e^{+}$ for SM and signal with different anomalous coupling $\kappa$ values at $\sqrt{s}=0.5,1.5$ and $3$ TeV.
\label{fig7}}
\end{figure}

\begin{table}
\caption{Parameters of the CLIC in three different stages. Here $\sqrt{s}$ is the center-of-mass energy, $L$ is the total luminosity, $N$ is the number of particles in bunch, $\sigma_{x}$
and $\sigma_{y}$ are the beam sizes and $\sigma_{z}$ is the bunch length \cite{CLIC1}.
\label{tab1}}
\begin{ruledtabular}
\begin{tabular}{ccccc}
Parameter& Unit& Stage$1$& Stage$2$& Stage$3$ \\
\hline
$\sqrt{s}$& TeV& $0.5$& $1.5$& $3$ \\
$L$& $10^{34}$ cm$^{-2}$ s$^{-1}$& $2.3$& $3.2$& $5.9$ \\
$N$& $10^{9}$& $3.7$& $3.7$& $3.7$ \\
$\sigma_{x}$& nm& $100$& $60$& $40$ \\
$\sigma_{y}$& nm& $2.6$& $1.5$& $1$  \\
$\sigma_{z}$& $\mu$m& $44$& $44$& $44$  \\
\end{tabular}
\end{ruledtabular}
\end{table}

\begin{table}
\caption{95\% C.L. sensitivity bounds of anomalous $\kappa_{tq\gamma}$ coupling
for various integrated CLIC luminosities and virtualities of the photon at the $\sqrt{s}=0.5$ TeV.
\label{tab2}}
\begin{ruledtabular}
\begin{tabular}{cccc}
Luminosity($fb^{-1}$)& $Q_{max}^2=4$\, GeV$^2$& $Q_{max}^2=16$\, GeV$^2$&  $Q_{max}^2=64$\, GeV$^2$  \\
\hline
$10$& $0.318$& $0.291$& $0.271$ \\
$50$& $0.138$& $0.126$& $0.116$  \\
$100$& $0.094$& $0.11$& $0.101$  \\
$230$& $0.074$& $0.067$& $0.061$ \\
\end{tabular}
\end{ruledtabular}
\end{table}

\begin{table}
\caption{95\% C.L. sensitivity bounds of anomalous $\kappa_{tq\gamma}$ coupling
for various integrated CLIC luminosities and virtualities of the photon at the $\sqrt{s}=1.5$ TeV.
\label{tab3}}
\begin{ruledtabular}
\begin{tabular}{cccc}
Luminosity($fb^{-1}$)& $Q_{max}^2=4$\, GeV$^2$& $Q_{max}^2=16$\, GeV$^2$&  $Q_{max}^2=64$\, GeV$^2$  \\
\hline
$10$& $0.101$& $0.0093$& $0.088$ \\
$100$& $0.043$& $0.038$& $0.038$  \\
$200$& $0.033$& $0.032$& $0.031$  \\
$320$& $0.026$& $0.026$& $0.025$ \\
\end{tabular}
\end{ruledtabular}
\end{table}

\begin{table}
\caption{95\% C.L. sensitivity bounds of anomalous $\kappa_{tq\gamma}$ coupling
for various integrated CLIC luminosities and virtualities of the photon at the $\sqrt{s}=3$ TeV.
\label{tab4}}
\begin{ruledtabular}
\begin{tabular}{cccc}
Luminosity($fb^{-1}$)& $Q_{max}^2=4$\, GeV$^2$& $Q_{max}^2=16$\, GeV$^2$&  $Q_{max}^2=64$\, GeV$^2$  \\
\hline
$10$& $0.085$& $0.078$& $0.054$ \\
$100$& $0.032$& $0.028$& $0.027$  \\
$300$& $0.021$& $0.021$& $0.021$  \\
$590$& $0.018$& $0.018$& $0.018$ \\
\end{tabular}
\end{ruledtabular}
\end{table}

\end{document}